\documentclass{PoS}
\usepackage{amsmath,amssymb,epsfig,fleqn,url,color,here,graphics,graphicx,rotating,multirow,wrapfig,enumerate}
\usepackage{etex}
\usepackage{caption}
\usepackage{subcaption}
\usepackage{pbox}
\usepackage{dcolumn}
\usepackage{bigstrut}
\usepackage{soul,xcolor}
\usepackage{lineno, xcolor}

\usepackage[section]{placeins}

\newcommand{\lf}{\left}
\newcommand{\rg}{\right}
\newcommand{\phiCS}{\varphi _{CS}}
\newcommand{\phiS}{\varphi _S}
\newcommand{\thCS}{\theta _{CS}}
\newcommand{\gvc}{GeV/$c$}

\newcommand{\gvcw}{GeV/$c^2$}

\newcommand{\Minv}{$M_{\mu\mu}$}

\setstcolor{red}
\title{Transversely polarized Drell-Yan measurements at COMPASS}

\ShortTitle{Transversely polarized Drell-Yan measurements at COMPASS}

\author{\speaker{Bakur Parsamyan} \thanks{On Behalf of the COMPASS collaboration}\\
        CERN, University of Turin and INFN section of Turin\\
        E-mail: \email{bakur.parsamyan@cern.ch}}

\abstract{The exploration of the transverse spin structure of the nucleon by measuring spin (in)dependent azimuthal asymmetries in semi-inclusive DIS (SIDIS) and in Drell-Yan processes is one of the main objectives of the COMPASS experiment at CERN (SPS, M2 beamline). During the first phase of the experiment (2002-2011) a series of SIDIS measurements were performed, using a longitudinally polarized muon beam impinging on transversely polarized $^6$LiD or NH$_3$ targets. As a part of the COMPASS-II programme, in 2015 and 2018 the experiment performed Drell-Yan measurements with a $\pi^-$ beam interacting with a transversely polarized NH$_3$. The measurement of the Sivers and other azimuthal asymmetries at the same hard scale in polarized SIDIS and Drell-Yan provides a unique possibility to test predicted in QCD (pseudo-)universal features of transverse momentum dependent parton distribution functions.}

\FullConference{XXVII International Workshop on Deep-Inelastic Scattering and Related Subjects - DIS2019\\
		8-12 April, 2019\\
		Torino, Italy}
\begin{document}
\section{Introduction}
\label{sec:intro}
The TMD PDFs are universal, process-independent functions\footnote{QCD \textit{generalized universality}: time-reversal modified process-independence of TMD PDFs}~\cite{Collins:2011zzd} describing longitudinal and transverse momenta distributions of partons and their correlations with nucleon and quark spins. Within Leading Order (LO) QCD parton model approach the polarized nucleon is described by six time reversal even and two time reversal odd \textit{twist-2} quark transverse momentum dependent (TMD) parton distribution functions (PDFs). Correlations between nucleon and quark spins and quark intrinsic momenta induce azimuthal modulations (asymmetries) in the cross sections of SIDIS ($\ell \,N \rightarrow \ell^\prime \,h \, X$, semi-inclusive hadron production in deep-inelastic lepton-nucleon scattering) and of Drell-Yan (DY) process ($h \, N \rightarrow \ell\,\bar{\ell}\, X$, massive lepton-pair production in hadron-nucleon collisions). Applying the TMD factorization theorems~\cite{Collins:2011zzd} allows one to express the asymmetries arising in DY and SIDIS cross sections in terms of convolutions of perturbatively calculable hard-scattering parton cross sections, hard-scale dependent TMD PDFs and (for SIDIS) parton fragmentation functions (FFs). The hard scale $Q$ in SIDIS is given by the square root of the virtuality of the photon exchanged in the DIS process and in DY by the invariant mass of the produced lepton pair.

%Measurements and following study of the spin (in)dependent azimuthal effects in SIDIS and Drell-Yan is a powerful method to access TMD distribution functions of the nucleon, which in past decades became a priority direction in experimental and theoretical high-energy physics, for recent reviews see \textit{e.g.} Refs.~\cite{Aidala:2012mv,Boglione:2015zyc}.
%
%The ultimate goal is to measure experimentally with high precision all possible spin-effects with both SIDIS and Drell-Yan reactions at different energies and perform global multi-differential analysis of obtained results to extract all spin-dependent distribution functions.
%%

\begin{wrapfigure}{r}{6.5cm}
\centering
\includegraphics[width=0.4\textwidth]{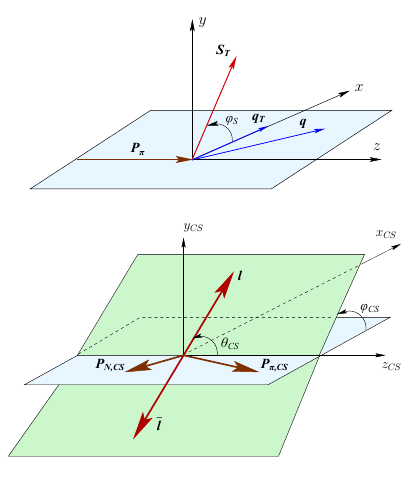}
\caption{The target rest (top) and the Collins-Soper frames (bottom).
}
\label{fig:DYframe}
\end{wrapfigure}

When the polarizations of the produced leptons are summed over, the general expression for the cross section of pion induced DY lepton-pair production off a transversely polarized nucleon comprises one polar asymmetry, two \textit{unpolarized} and five target transverse-spin-dependent azimuthal asymmetries (TSAs). Adopting general notations and conventions of Refs.~\cite{Arnold:2008kf,Gautheron:2010wva}, the model-independent differential cross section can be written as follows:
{\small
\begin{eqnarray}\label{eq:DY_xsecLO}
  \hspace*{0cm}\frac{d\sigma}{dq^{4}d\Omega} &\propto& \hat{\sigma}_{U}\bigg\{ 1 + D_{[\sin{2\thCS}]} A_U^{\cos {\phiCS}}\cos {\phiCS}+
  D_{[\sin^{2}\thCS]} A_U^{\cos 2{\phiCS}}\cos 2{\phiCS}\\ \nonumber
   &&\hspace*{0.9cm} +  {S_T}\Big[ D_{[1+\cos^{2}\thCS]}A_T^{\sin {\phiS}}\sin {\phiS} \\ \nonumber
   &&\hspace*{1.4cm} + D_{[\sin^{2}\thCS]}
       \lf( A_T^{\sin( {2{\phiCS} - {\phiS}}    )}\sin ( {2{\phiCS} - {\phiS}} )
   { + A_T^{\sin \lf( {2{\phiCS} + {\phiS}} \rg)}}\sin \lf( {2{\phiCS} + {\phiS}} \rg) \rg)\\ \nonumber
   &&\hspace*{1.4cm} + D_{[\sin{2\thCS}]}
       \lf( A_T^{\sin( {{\phiCS} - {\phiS}}    )}\sin ( {{\phiCS} - {\phiS}} )
   { + A_T^{\sin \lf( {{\phiCS} + {\phiS}} \rg)}}\sin \lf( {{\phiCS} + {\phiS}} \rg) \rg)
   \Big] \bigg\},
\end{eqnarray}
}
Here, $q$ is the four-momentum of the exchanged virtual photon, $F^{1}_{U}$, $F^{2}_{U}$ are the polarization and azimuth-independent structure functions and $\hat{\sigma}_{U} =  \lf({F^{1}_{U}}+{F^{2}_{U}}\rg)\lf(1 + \lambda{{\cos}^2}\thCS \rg)$, with $\lambda$ being the polar angle asymmetry, given as $\lambda=\lf({F^{1}_{U}}-{F^{2}_{U}}\rg)/\lf({F^{1}_{U}}+{F^{2}_{U}}\rg)$. The subscript ($U$)$T$ denotes target transverse polarization (in)dependence. In analogy to SIDIS, the virtual-photon depolarization factors are defined as $D_{[f(\thCS)]}=f(\thCS)/\lf(1 + \lambda{{\cos }^2}\thCS \rg)$ with $f(\thCS)$ being equal either to $\sin2\thCS$, or $\sin^2\thCS$, or to $1+\cos^2\thCS$. The angles $\phiCS$, $\thCS$ and $\Omega$, the solid angle of the lepton, are defined in the Collins-Soper frame following the considerations of Refs.~\cite{Arnold:2008kf,Gautheron:2010wva}, and $\phiS$ is the azimuthal angle of the direction of the nucleon polarization in the target rest frame, see Fig.~\ref{fig:DYframe}.

The asymmetries $A_{(U)T}^{w}$ in Eq.~\ref{eq:DY_xsecLO} are defined as amplitudes of a given azimuthal modulation $w=w(\phiS,\phiCS)$, divided by the spin and azimuth-independent part of the DY cross section and the corresponding depolarization factor.

%In this analysis, the sign convention for TSAs is given by Eq.~(\ref{eq:DY_xsecLO}) together with the definitions of azimuthal and polar angles in Fig.~\ref{fig:DYframe}.
One spin-independent asymmetry and three out of five TSAs enter at leading order of  perturbative QCD and can be described by contributions from only \textit{twist-2} TMD PDFs. In DY lepton-pair production with a transversely polarized nucleon in the initial state, the $A_U^{\cos2\phiCS}$ asymmetry is related to the convolution of nucleon and pion Boer-Mulders TMD PDFs, ($h_{1}^{\perp}$ and $h_{1,\pi}^{\perp}$, correspondingly).
The TSA $A_T^{\sin\phiS}$ is related to the nucleon Sivers TMD PDFs ($f_{1T}^\perp$) convoluted with the unpolarized pion TMD PDFs ($f_{1,\pi}$). Here, following similar SIDIS conventions, the \textit{twist-3} contribution to the Sivers TSA is neglected\footnote{the $F^{2}_{UT}$ structure function is assumed to be zero~\cite{Gautheron:2010wva}}. Analogously, within the \textit{twist-2} approximation of LO pQCD, $F^{2}_{U}=0$ and therefore $\lambda=1$.
The other two \textit{twist-2} TSAs, $A_T^{\sin(2\phiCS-\phiS)}$ and $A_T^{\sin(2\phiCS+\phiS)}$, are related to convolutions of the Boer-Mulders TMD PDFs ($h_{1,\pi}^{\perp}$) of the pion with the nucleon TMD PDFs transversity ($h_1$) and pretzelosity ($h_{1T}^\perp$), respectively~\cite{Arnold:2008kf,Bacchetta:2006tn}. The remaining three azimuthal asymmetries, namely $A_U^{\cos\phiCS}$, $A_T^{\sin(\phiCS-\phiS)}$ and $A_T^{\sin(\phiCS+\phiS)}$ are \textit{subleading-twist} structures and are expected to vanish at LO.

All three aforementioned \textit{twist-2} nucleon TMD PDFs related to LO DY TSAs induce analogous \textit{twist-2} TSAs in the general expression for the cross section of unpolarized-hadron production in SIDIS of leptons off transversely polarized nucleons~\cite{Arnold:2008kf,Bacchetta:2006tn,Kotzinian:1994dv}.
The SIDIS TSAs were measured by the HERMES and COMPASS experiments, see Refs.~\cite{Adolph:2012sp,Adolph:2012sn,Airapetian:2009ae,Adolph:2016dvl,Parsamyan:2013ug}.

The Sivers function~\cite{Sivers:1989cc} plays an important role among the TMD PDFs.
In the TMD framework of QCD it is predicted that the two naively time-reversal odd TMD PDFs, \textit{i.e.} the quark Sivers functions $f_{1T}^\perp$ and Boer-Mulders functions $h_{1}^\perp$, have opposite sign when measured in SIDIS and in DY (or $W$/$Z$-boson) productions.
%, which is the consequence of different Wilson-line structures in the two processes
~\cite{Collins:2002kn}.
%non-pertutbative QCD object PDF gauge invraiance
The experimental test of this fundamental prediction, which is a direct consequence of QCD gauge invariance, is a major challenge in hadron physics. In contrast to the  Sivers and Boer-Mulders functions, transversity and pretzelosity TMD PDFs are predicted to be genuinely universal, \textit{i.e.} they do not change sign between SIDIS and DY~\cite{Collins:2011zzd}, which is yet another fundamental QCD prediction to be explored.

Non-zero quark Sivers TMD PDFs have been extracted from SIDIS single-differential results of HERMES~\cite{Airapetian:2009ae}, COMPASS~\cite{Adolph:2012sp} and JLab~\cite{Qian:2011py} using both collinear and TMD evolution approaches~\cite{Anselmino:2016uie,Echevarria:2014xaa,Sun:2013hua}.
The first measurement of the Sivers effect in $W$ and $Z$-boson production in collisions of transversely polarized protons at RHIC was reported by the STAR collaboration~\cite{Adamczyk:2015gyk};
the hard scales of these measurements is $Q\approx 80$\;\gvc\; and $90$\;\gvc. It is quite different from the one explored in fixed-target experiments and it is not excluded that TMD evolution effects may be sizable when describing the STAR results using Sivers TMD PDFs extracted from fixed-target SIDIS results.

%%
%\begin{wrapfigure}{r}{5cm}
%\centering
%\includegraphics[width=0.33\textwidth]{plots/A8in4Q2range_Zgt01_Pr10.pdf}
%\caption{COMPASS results for proton SIDIS TSAs in the range 4 \gvc$\;<Q<$ 9 \gvc.}
%\label{fig:A8in1}
%\end{wrapfigure}
%%
The COMPASS experiment at CERN~\cite{Gautheron:2010wva} has the unique capability to explore the transverse-spin structure of the nucleon in a similar kinematic region by two alternative experimental approaches, \textit{i.e.} SIDIS and DY.
This offers the opportunity of minimizing the uncertainties related to the TMD evolution in the comparison of the Sivers TMD PDFs when extracted from these two measurements to test the opposite-sign prediction by QCD.
Recently, COMPASS published the first multi-differential results of the TSAs, which were extracted from SIDIS data at four different hard scales~\cite{Adolph:2016dvl}. In order to simplify the comparison with the Drell-Yan case, one of the selected $Q^2$ ranges was chosen to be the same as the one selected for DY TSAs analysis.
%The three aforementioned TSAs that are extracted from COMPASS SIDIS data for the range 4 \gvc$\;<Q<$ 9 \gvc, are shown in Fig.~\ref{fig:A8in1} after averaging over all other kinematic dependences. This hard scale range is very similar to the one used in this Letter to analyze the DY process.
%The Sivers asymmetry for positive hadrons was found to be above zero by 3.2 standard deviations of the total experimental accuracy. The Collins asymmetry is determined with even better statistical precision. The amplitude has opposite sign for positive and negative hadrons which is attributed to the peculiarities of Collins FF~\cite{Adolph:2012sn}. The pretzelosity asymmetry in SIDIS is found to be compatible with zero, which can be related to kinematic suppression-factors entering in the corresponding structure function~\cite{Bacchetta:2006tn,Parsamyan:2013ug}.
%
%

COMPASS results for three \textit{twist-2} Drell-Yan TSAs from 2015 measurements were published in Ref.~\cite{Aghasyan:2017jop}. In this Letter preliminary COMPASS results obtained from the part ($\approx$50\%) of the 2018 data are shown in combination with 2015 TSA measurements.

\section{Data analysis}
\label{sec:data_analysis}
\begin{wrapfigure}{r}{6.5cm}
\centering
\includegraphics[width=0.4\textwidth]{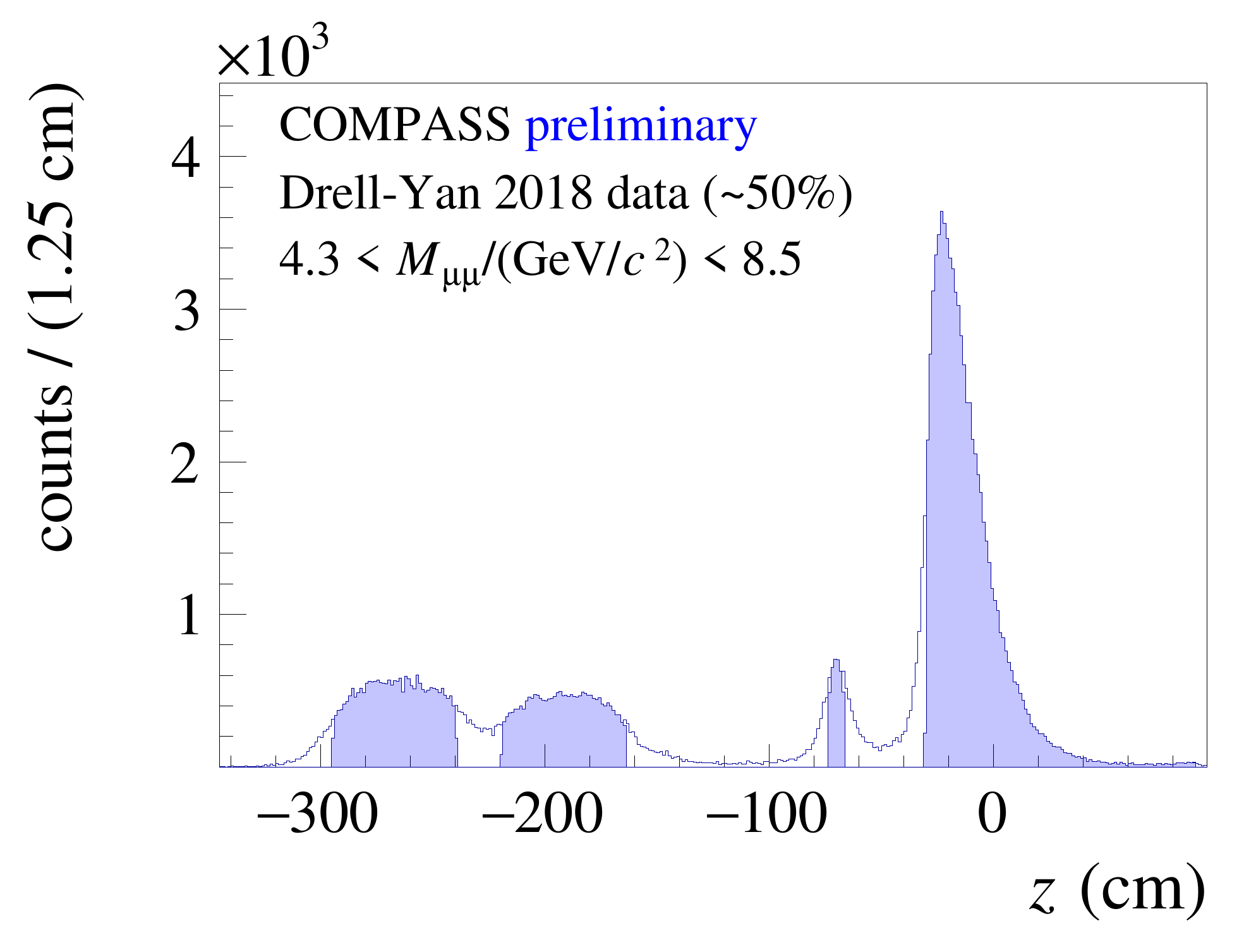}
\caption{Distribution of dimuon production vertices in the target region.}
\label{fig:PVz}
\end{wrapfigure}

The analysis presented in this Letter is based on Drell-Yan data collected by COMPASS in the years 2015 and 2018. So far only part of 2018 data ($\approx$50\%) has been processed and analyzed.
For the measurement, the 190 \gvc\; $\pi^{-}$ beam with an average intensity of $0.6\times10^{8}$ s$^{-1}$ was scattered off the COMPASS transversely polarized NH$_3$ target.
The polarized target, placed in a 0.6~T dipole magnet, consisted of two longitudinally aligned cylindrical cells of 55 cm length and 4~cm in diameter, separated
by a 20~cm gap.
A 240 cm long structure made mostly of aluminum oxyde with a tungsten core, placed downstream of the target, acted as hadron absorber and beam dump. %In addition, a 7cm long aluminum plug dedicated to the unpolarized DY-measurements was installed inside the absorber upstream of the tungsten beam plug. %Approximately $60\%$ of the beam did not interact in the polarized target and was dumped into the tungsten plug.
The distribution of dimuon production vertices in the target region is shown in Fig.~\ref{fig:PVz} with indicated target cell, aluminum target and tungsten core positions.
%
%The average uncertainty in the reconstructed vertex position along the beam line is about $10$~cm, which leads to a marginal (below 1\%)  migration of reconstructed events from one target cell to the other.
%
The two cells were polarized vertically in opposite directions, so that data with both spin orientations were recorded simultaneously. In order to compensate for acceptance effects, the polarization was reversed approximately every two weeks.
%
%The entire data-taking time of 18 weeks was divided into nine periods, each consisting of two consecutive weeks with opposite target polarizations.
%The proton polarization had a relaxation time of about 1000 hours, which was measured for each target cell in each data taking period.
The average proton polarization was measured to be $\langle P_T\rangle\approx0.73$. The dilution factor, $f$, accounting for the fraction of polarizable nucleons in the target and the migration of reconstructed events from one target cell to the other or from unpolarized medium into the cells, is estimated to be $\langle f\rangle\approx 0.18$.

%
%
%\begin{wrapfigure}{r}{5.5cm}
%\centering
%\includegraphics[width=0.35\textwidth]{plots/f_15.pdf}
%\caption{Average dilution factor in bins of $x_{N}$ and $x_{\pi}$.}
%\label{fig:f_x}
%\end{wrapfigure}
%%

%Outgoing charged particles were detected by a system of tracking detectors in the two-stage spectrometer and their momenta are determined by means of two large-aperture dipole magnets. In each stage, muon identification was accomplished by a system of muon filters.
%
%The trigger required the hit pattern of
%several hodoscope planes to be consistent with at least two muon candidates originating from the target region. For any pair of candidates either both have to be detected in the first stage of the spectrometer ($25 < \theta_{\mu} < 160$~mrad), or one in the first and the other in the second stage ($8 < \theta_{\mu} < 45$~mrad).

\begin{figure}[h!]
\centering
\begin{subfigure}{0.35\textwidth}
\includegraphics[width=1.0\textwidth]{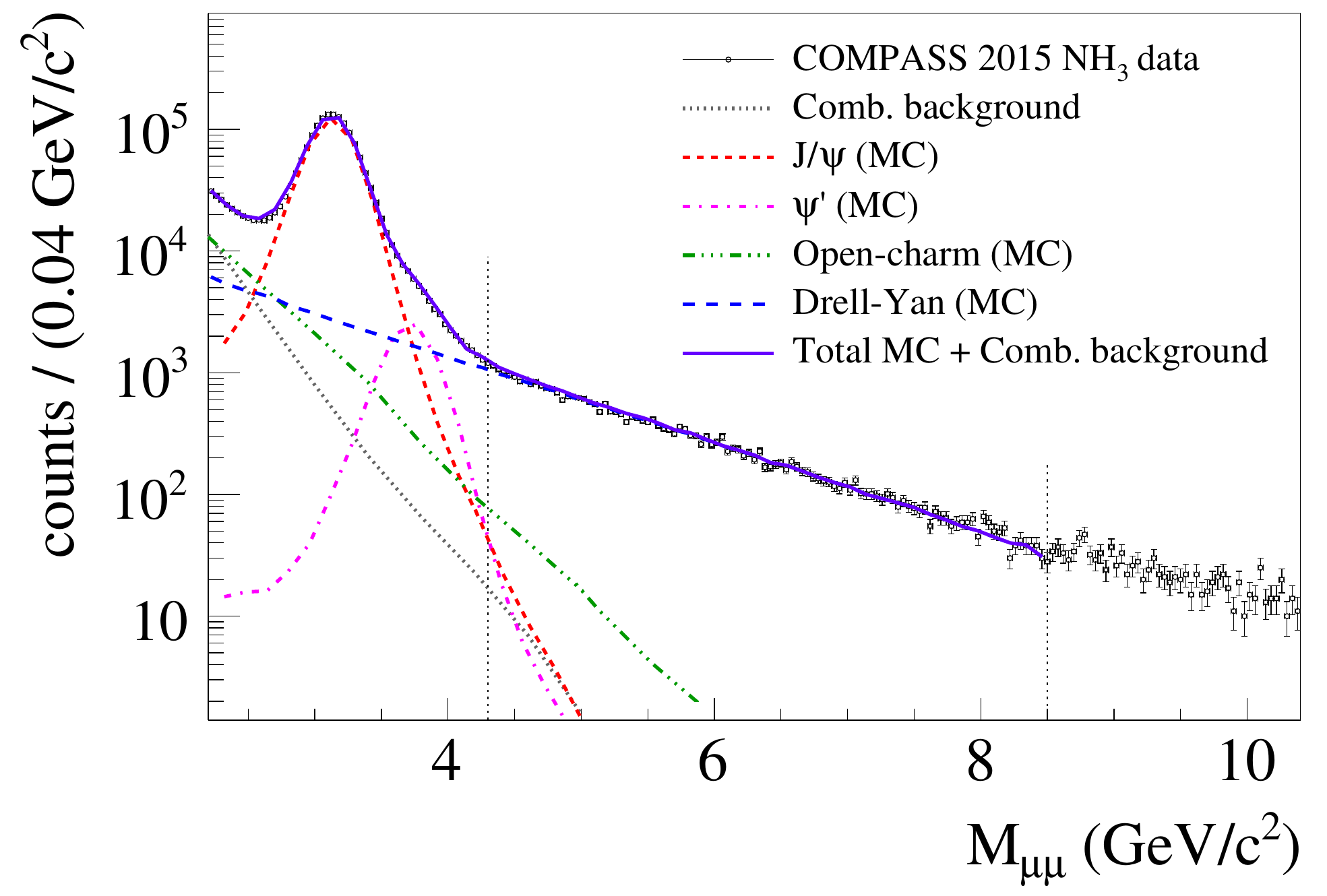}
\includegraphics[width=1.0\textwidth]{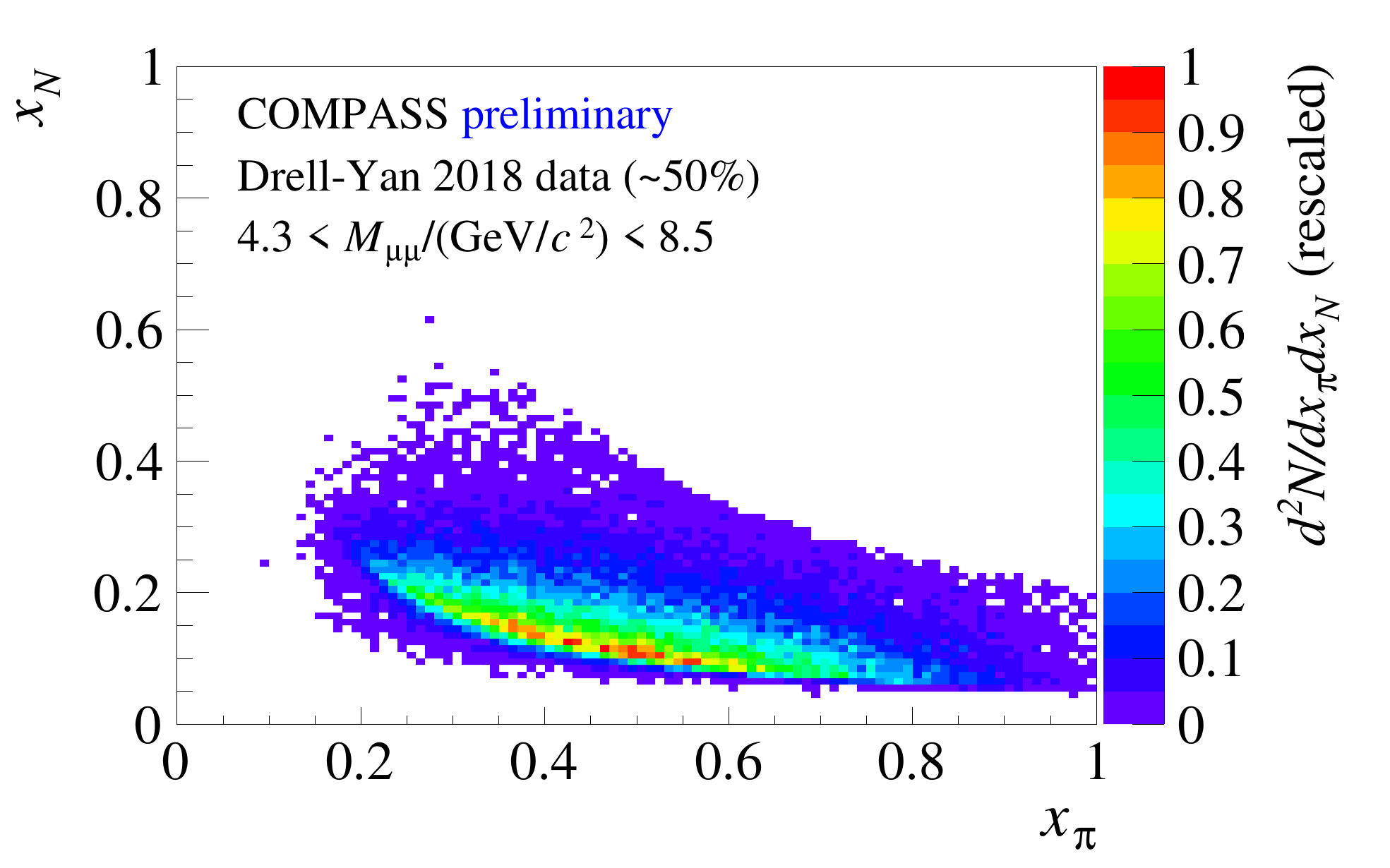}
\end{subfigure}
\hspace*{0.5cm}
\begin{subfigure}{0.6\textwidth}
\includegraphics[width=1.0\textwidth]{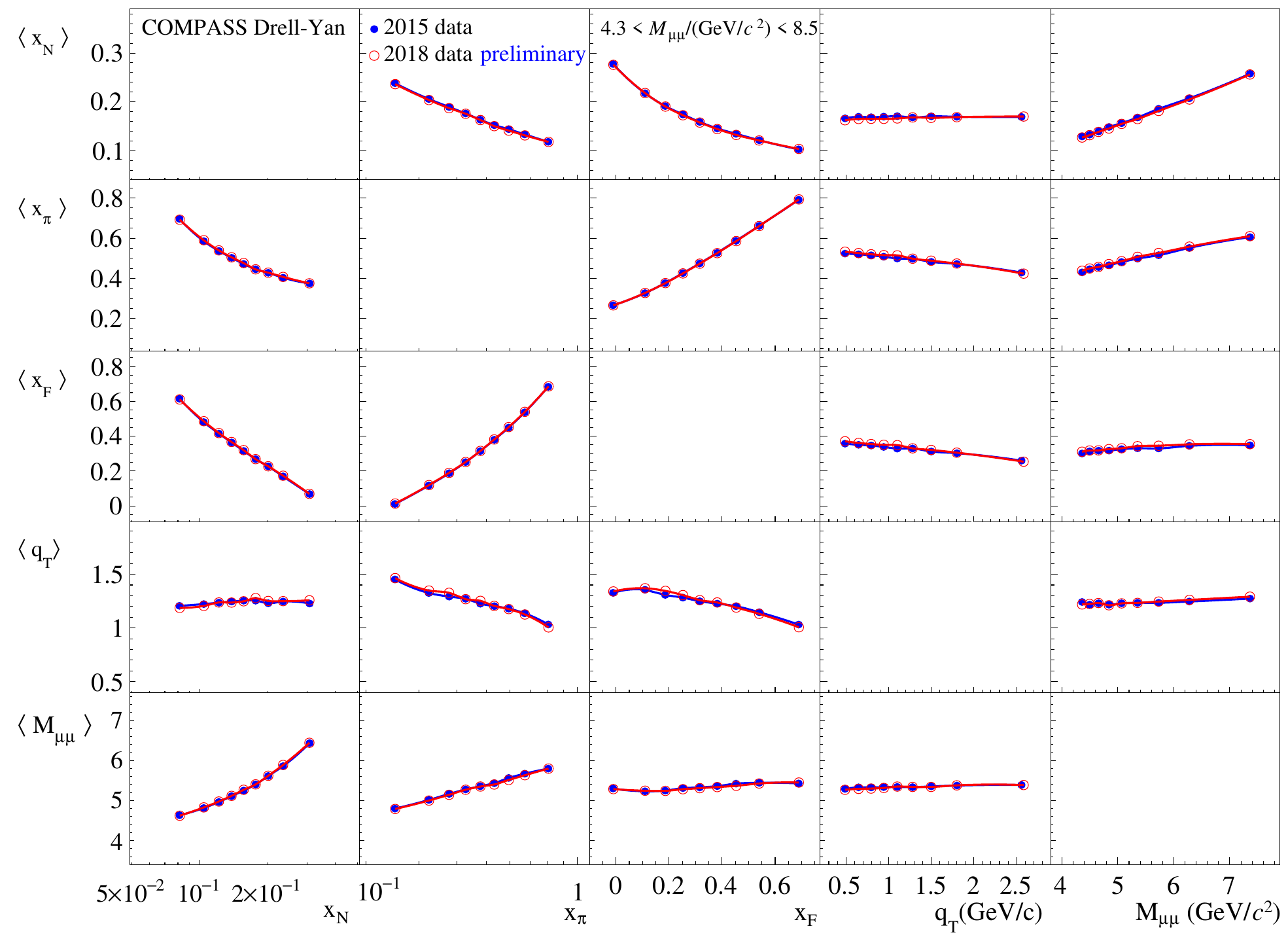}
\end{subfigure}%
\caption{Left top panel: the \Minv distribution. Left bottom panel: the two-dimensional $(x_{\pi},x_{N})$ distribution of the selected high mass dimuons. Right panel: dependences of average kinematic values as extracted from 2015 and 2018 data.}
\label{fig:kine}
\end{figure}

%
%In the data analysis, the selection of events requires a production vertex located within the polarized-target volume, with one incoming pion beam track and at least two oppositely charged outgoing particles that are consistent with the muon hypothesis, ensured by the requirement of 30 radiation lengths to be crossed along the spectrometer. The quality and timing information for the tracks is also verified.
%
The dimuon invariant mass range selected for TSA analysis is defined by the requirement 4.3~\gvcw\,$ < M_{\mu\mu} < $\,8.5~\gvcw\,, where the upper limit avoids the contribution of $\Upsilon$-resonances. In this range, the overall background contribution to the Drell-Yan sample is estimated to be below 4\%.
The reconstructed mass spectrum of events passing all analysis requirements is shown in black in Fig.~\ref{fig:kine} (left top panel).
The two-dimensional distribution of the Bjorken scaling variables of pion and nucleon, $x_{\pi}$ and $x_{N}$, for this mass range is presented in Fig.~\ref{fig:kine} (left bottom panel). The figure demonstrates that the kinematic phase space explored by the COMPASS spectrometer matches the valence region in $x_{\pi}$ and $x_{N}$. In this region, the DY cross section for a proton target is dominated by the contribution of nucleon $u$-quark and pion $\bar{u}$-quark TMD PDFs.

%%
%\begin{wrapfigure}{r}{6.5cm}
%\centering
%\includegraphics[width=0.4\textwidth]{plots/DY_MassFit.pdf}
%\includegraphics[width=0.4\textwidth]{plots/DY2018_Mr4_NH3_xbxN.pdf}
%\caption{Yop panel: the \Minv distribution. Bottom panel: the two-dimensional $(x_{\pi},x_{N})$ distribution of the selected high mass dimuons}
%\label{fig:M}
%\end{wrapfigure}
%%

The dimuon transverse momentum $q_T$ is required to be above $0.4$ \gvc\; in order to obtain sufficient resolution in angular variables.
The comparison of kinematic dependences of average values of different kinematic variables between 2015 and 2018 data is shown in Fig.~\ref{fig:kine} (left panel). The kinematic dependences in two years are nearly identical.

%%
%\begin{wrapfigure}{r}{5.5cm}
%\centering
%\includegraphics[width=0.35\textwidth]{plots/DY2015_NH3_xNQ2_Mranges.pdf}
%\caption{the two-dimensional $(x_{N},Q^2)$ distribution.}
%\label{fig:xQ2}
%\end{wrapfigure}
%%

%The distributions of the dimuon Feynman variable $x_{F}$ and the dimuon transverse momentum $q_{T}$ are presented in Fig.~\ref{fig:xqT} (central and right panels). The corresponding mean values of the kinematic variables are: $\langle x_{N} \rangle=0.17$,
%$\langle x_{\pi} \rangle=0.50$,
%$\langle x_{F} \rangle=0.33$,
%$\langle q_{T} \rangle=1.2$~\gvc\, and
%$\langle M_{\mu\mu} \rangle=5.3$~\gvcw.

%After all selection criteria about $35\times10^3$ dimuons remain for the analysis.
%The TSAs presented in this Letter are extracted period by period from the number of dimuons produced in each cell for the two directions of the target polarization.
%The double-cell target configuration in conjunction with the periodic polarization reversal allows for the simultaneous measurement of azimuthal asymmetries for both target spin orientations.
%

%
Using an extended Unbinned Maximum Likelihood estimator, all five TSAs are fitted simultaneously together with their correlation matrices. %In this approach, flux and acceptance-dependent systematic uncertainties are minimized~\cite{Adolph:2016dvl}.
%The final asymmetries are obtained by averaging the results of the nine periods.
The asymmetries are evaluated in kinematic bins of $x_{N}$, $x_{\pi}$, $x_{F}$ or $q_{T}$, while always integrating over all the other variables.
%
%\begin{wrapfigure}{r}{5.5cm}
%%\begin{figure}[h!]
%\centering
%\includegraphics[width=0.35\textwidth]{plots/Dth_15.pdf}
%\caption{Kinematic dependence of the depolarization factors and impact of different values of $\lambda$.}
%\label{fig:Dth}
%\end{wrapfigure}
%%\end{figure}
%
%The dilution factor $f$ and the depolarization factors entering the definition of TSAs are calculated on an event-by-event basis and used to weight the asymmetries. For the magnitude of the target polarization $P_T$, an average value is used for each data taking period in order to avoid possible systematic bias. In the evaluation of the depolarization factors, the approximation $\lambda=1$ is used. Known deviations from this assumption with $\lambda$ ranging between 0.5 and 1~\cite{Falciano:1986wk} decrease the normalization factor by at most $5\%$, see Fig.~\ref{fig:Dth}.

%The TSAs resulting from different periods are checked for possible systematic effects. The largest systematic uncertainty is due to possible residual variations of experimental conditions within a given period. They are quantified by evaluating various types of false asymmetries in a similar way as described in Refs.~\cite{Adolph:2012sp,Adolph:2012sn}.
The systematic point-to-point uncertainties are estimated to be about 0.7 times the statistical uncertainties and are indicated in the plots by color bands.
%The normalization uncertainties originating from the uncertainties on target polarization ($5\%$) and dilution factor ($8\%$) are not included in the quoted systematic uncertainties.

%
\section{Results and Discussion}
\label{sec:results}
\begin{figure}[h!]
\centering
\begin{subfigure}{0.7\textwidth}
\includegraphics[width=0.9\textwidth]{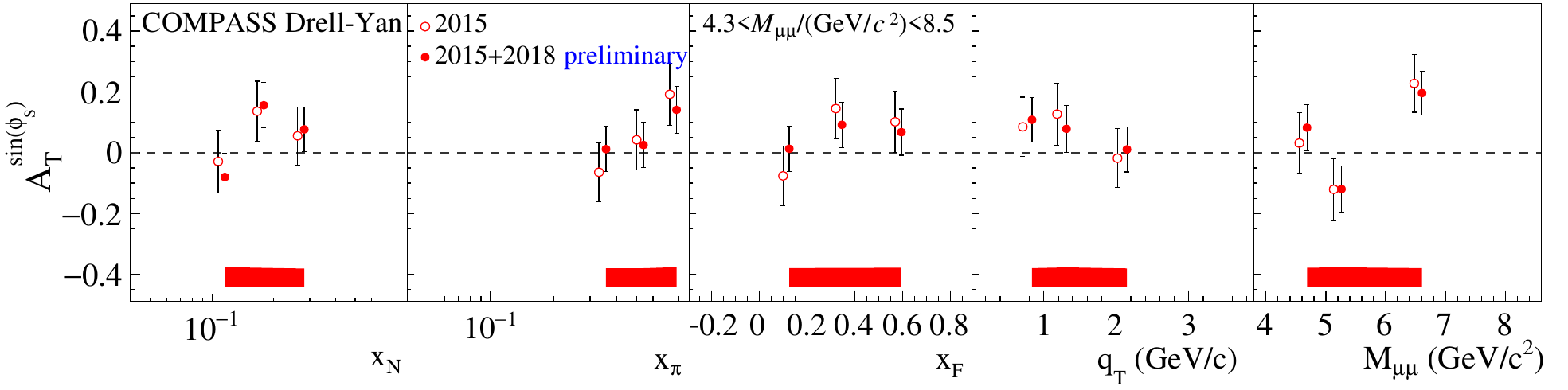}
\includegraphics[width=0.9\textwidth]{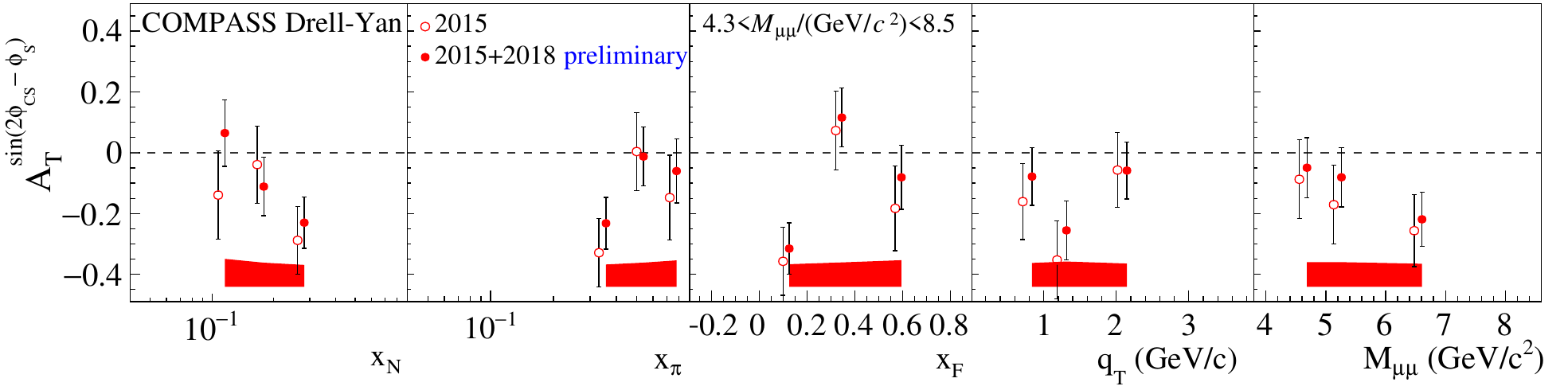}
\includegraphics[width=0.9\textwidth]{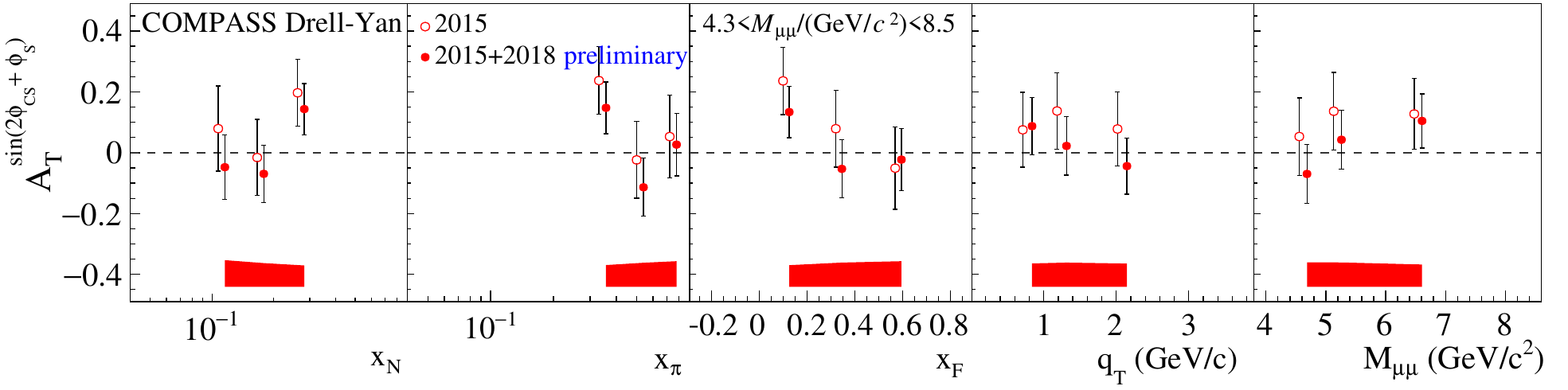}
\includegraphics[width=0.9\textwidth]{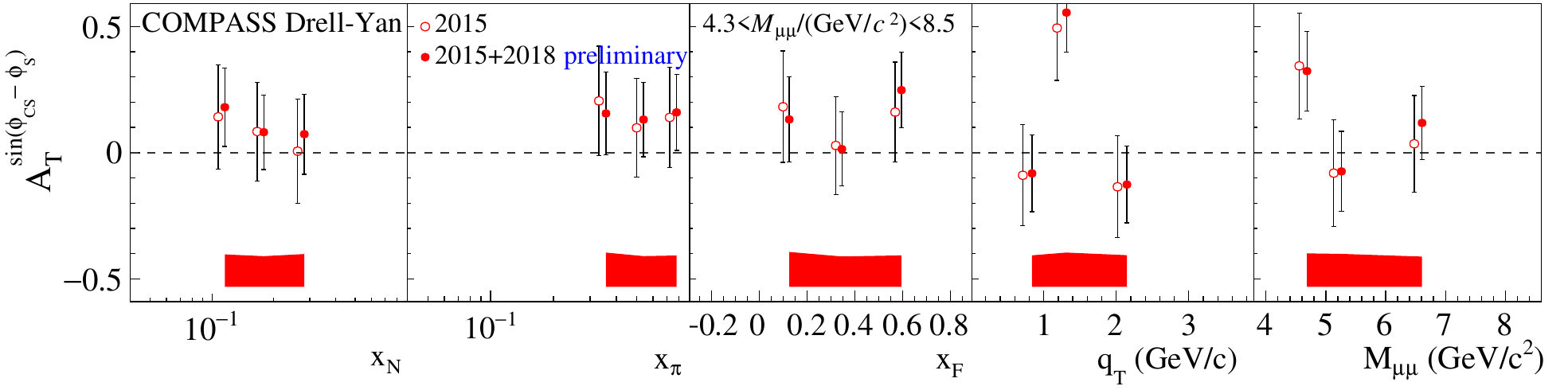}
\includegraphics[width=0.9\textwidth]{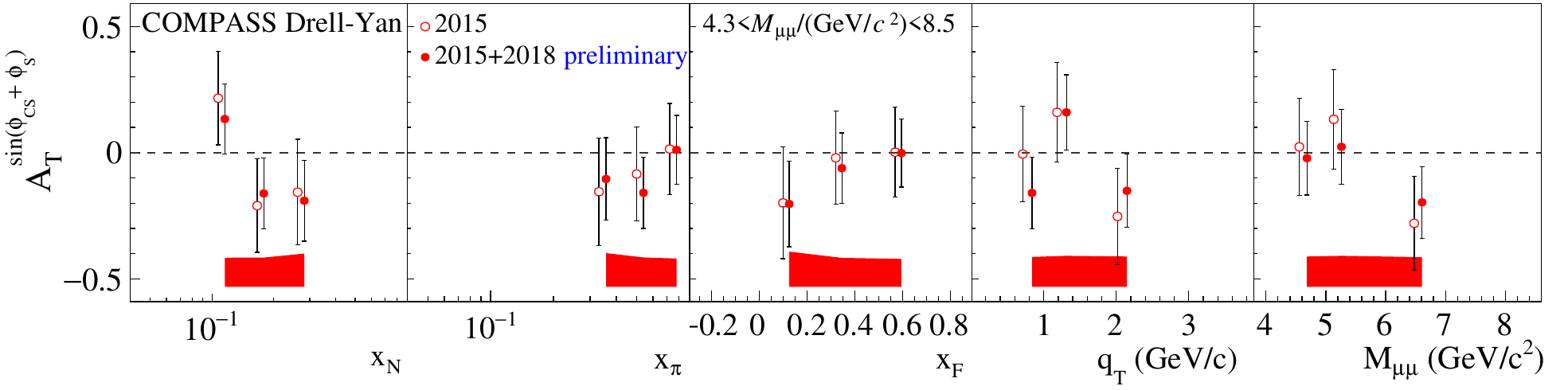}
\end{subfigure}%
\hspace*{-0.5cm}
\begin{subfigure}{0.3\textwidth}
\includegraphics[width=1.0\textwidth]{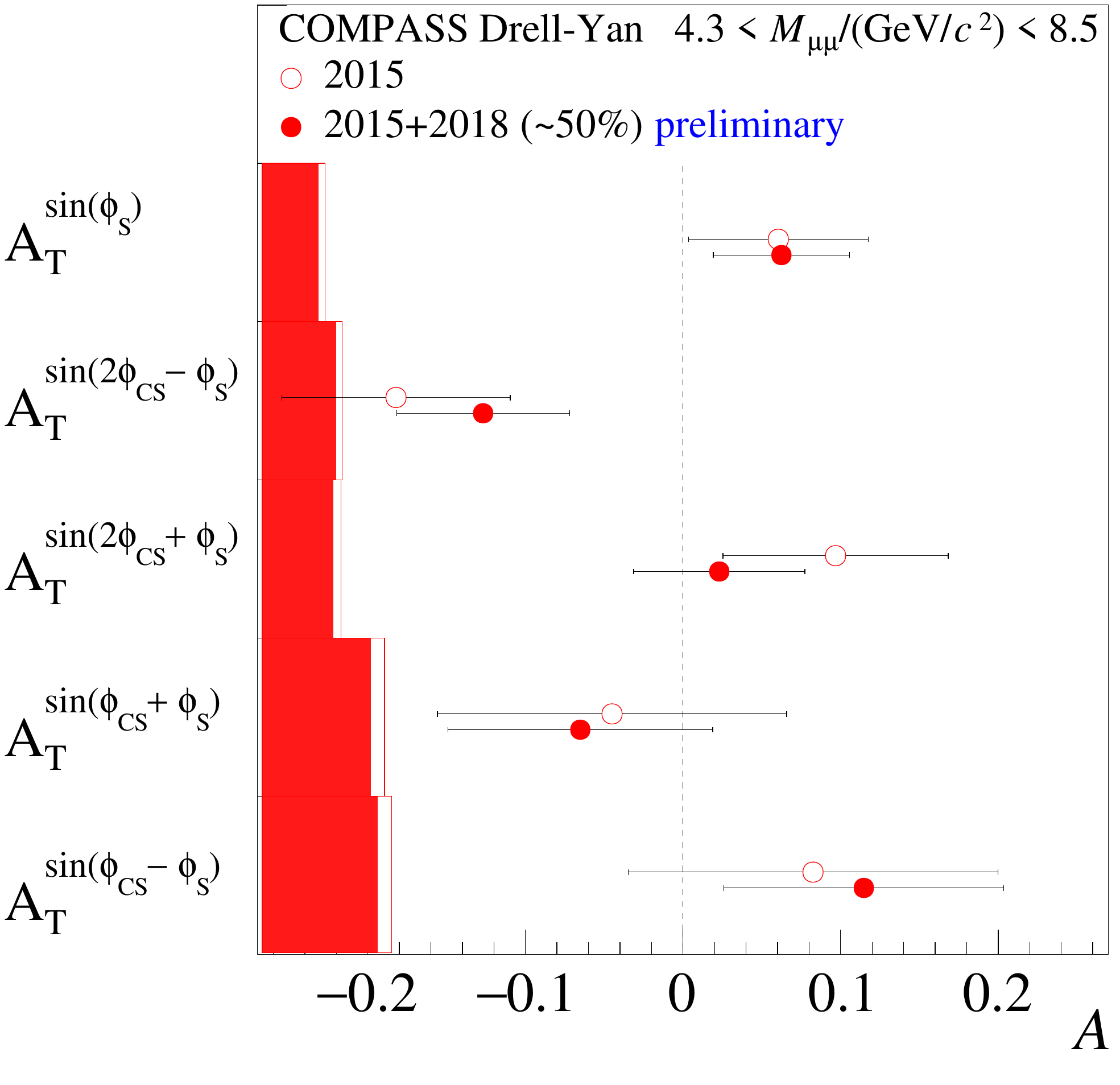}
\end{subfigure}
\caption{Left panel: Kinematic dependences of extracted Drell-Yan TSAs. Right panel: Drell-Yan TSAs integrated over the entire kinematic range.}
%{Extracted Drell-Yan TSAs related to Sivers, transversity and pretzelosity TMD PDFs (top to bottom). Inner (outer) error bars represent statistical (total experimental) uncertainties. The normalization uncertainties due to target polarization ($5\%$) and dilution factor ($8\%$) are not included in the error bars.}
\label{fig:TSAs}
\end{figure}

The kinematic dependences for all five Drell-Yan TSAs (\textit{twist-2} TSAs, $A_T^{\sin\phiS}$, $A_T^{\sin(2\phiCS-\phiS)}$ and $A_T^{\sin(2\phiCS+\phiS)}$ and \textit{subleading-twist} TSAs, $A_T^{\sin(\phiCS-\phiS)}$ and $A_T^{\sin(\phiCS+\phiS)}$) are shown in Fig.~\ref{fig:TSAs} (left panel).
Empty circles indicate results obtained from 2015 data, while filled circles correspond to combined 2015 and preliminary 2018 results.
Due to the relatively large statistical uncertainties, no clear trend is observed for any of the TSAs.

The rightmost panel in Fig.~\ref{fig:TSAs} shows the 2015 measurement and the combined result of 2015 and 2018 for the five extracted TSAs integrated over the entire kinematic range. The preliminary results obtained with part of the 2018 data are found to be in agreement with the published 2015 results~\cite{Aghasyan:2017jop}.
The combined result of 2015 and 2018 for the average Sivers asymmetry $A_T^{\sin\phiS}$ is found to be above zero at about one standard deviation of the total uncertainty.
%
%%\begin{figure}[h!]
%\begin{wrapfigure}{r}{6.0cm}
%\centering
%\includegraphics[width=0.36\textwidth]{plots/Sivers_xF.pdf}
%\caption{The measured mean Sivers asymmetry and the
%theoretical  predictions from Refs.~\cite{Anselmino:2016uie} (DGLAP),~\cite{Echevarria:2014xaa} (TMD1)
%and~\cite{Sun:2013hua} (TMD2).
%The dark-shaded (light-shaded) predictions are evaluated with (without) the sign-change hypothesis}.
%\label{fig:Siv_theor}
%\end{wrapfigure}
%%\end{figure}
%
%
The positive sign and the amplitude of the asymmetry is consistent with the predicted change of sign for the Sivers function.
The average value of the TSA  $A_T^{\sin(2\phiCS-\phiS)}$ is measured to be below zero with a significance of about two standard deviations. The obtained magnitude of the asymmetry is in agreement with the model calculations of Ref.~\cite{Sissakian:2010zza} and can be used to study the universality of the nucleon transversity function. The TSA $A_T^{\sin(2\phiCS+\phiS)}$, which is related to the nucleon pretzelosity TMD PDF, is found to be compatible with zero when combining the results from two years.
%Since both $A_T^{\sin(2\phiCS-\phiS)}$ and $A_T^{\sin(2\phiCS+\phiS)}$ are related to the pion Boer-Mulders PDFs, the obtained results may be used to study this function further and to possibly determine its sign.
%
%They may also be used to test the sign change of the nucleon Boer-Mulders TMD PDFs between SIDIS and DY as predicted by QCD~\cite{Collins:2002kn}, when combined with other past and future SIDIS and DY data related to target-spin-independent Boer-Mulders asymmetries~\cite{Falciano:1986wk,Airapetian:2012yg,Adolph:2014pwc}.
%
%
The remaining two asymmetries are the $A_T^{\sin(\phiCS-\phiS)}$ and $A_T^{\sin(\phiCS+\phiS)}$) \textit{subleading-twist} TSAs. Both amplitudes are found to be compatible with zero (see Fig.~\ref{fig:TSAs}), which can be attributed to the subleading nature of the effects and corresponding dynamic suppressions.
The presented results are the first and the only currently available data on the transverse-spin-dependent azimuthal asymmetries in the Drell-Yan process. COMPASS continues to analyze the data collected in 2018, which will improve the statistical precision of the Sivers and other azimuthal DY TSAs presented in this Letter.
%Combined with available SIDIS data, COMPASS DY results serve as a crucial input for study of universality of TMD PDFs and for general understanding of the transverse spin structure of the nucleon.

%
%

\bibliography{DIS2019_DY_biblio}{}

\end{document}